\documentclass[aps,prl,twocolumn,groupedaddress,longbibliography,nofootinbib]{revtex4-2}

\usepackage{graphicx}
\usepackage{amsmath}

\usepackage{color}
\usepackage{subcaption}

\newcommand{\equs}[2]{equations~(\ref{#1})-(\ref{#2})}

\newcommand{\fig}[1]{Fig.~\ref{#1}}

\renewcommand{\vec}[1]{\protect\mathbf{#1}}

\newcommand{\curl}{\vec{\boldsymbol{\nabla}}\mathbf{\times}}
\renewcommand{\div}{\vec{\boldsymbol{\nabla}}\cdot}
\newcommand{\f}[2]{\frac{#1}{#2}}
\newcommand{\dpart}[2]{\f{\partial #1}{\partial #2}}

\newcommand{\vA}{\vec{A}}

\newcommand{\vB}{\vec{B}}

\newcommand{\vE}{\vec{E}}

\newcommand{\vEMF}{\boldsymbol{\mathcal{E}}}
\newcommand{\mean}[1]{\overline{#1}}
\newcommand{\fluct}[1]{#1}

\newcommand{\fluctvE}{\fluct{\vec{e}}}
\newcommand{\fluctvA}{\fluct{\vec{a}}}
\newcommand{\fluctvB}{\fluct{\vec{b}}}
\newcommand{\fluctvU}{\fluct{\vec{u}}}

\newcommand{\meanB}{\mean{B}}

\newcommand{\meanvA}{\mean{\vA}}
\newcommand{\meanvB}{\mean{\vB}}

\newcommand{\meanvE}{\mean{\vE}}

\newcommand{\meanvEMF}{\mean{\vEMF}}

\begin{document}

\title{Helical turbulent nonlinear dynamo at large magnetic Reynolds numbers}

\author{F. Rincon}
\email[]{francois.rincon@irap.omp.eu}
\affiliation{CNRS, IRAP, 14 avenue Edouard Belin, F-31400 Toulouse, France}
\affiliation{Universit\'e de Toulouse, UPS-OMP, IRAP: Toulouse, France}

\date{\today}

\begin{abstract}
The excitation and further sustenance of 
large-scale magnetic fields in rotating astrophysical systems,
including planets, stars and galaxies, is generally thought to
involve a fluid magnetic dynamo effect driven by helical
magnetohydrodynamic turbulence. While this scenario is appealing on general
grounds, it however currently remains largely unconstrained,
notably because a fundamental understanding of the nonlinear asymptotic behaviour
of large-scale fluid magnetism in the astrophysically-relevant but
treacherous regime of large magnetic Reynolds number $Rm$ is still
lacking. We explore this problem using local high-resolution
simulations of turbulent magnetohydrodynamics driven by an inhomogeneous
helical forcing generating a sinusoidal profile of
kinetic helicity, mimicking the hemispheric distribution of kinetic helicity in
rotating turbulent fluid bodies. We identify a transition at large
$Rm$ to a nonlinear state, followed up to $Rm\simeq 3\times
10^3$, consisting of strong, saturated small-scale magnetohydrodynamic
turbulence and a weaker, travelling coherent large-scale field
oscillation. This state is characterized by an asymptotically small
resistive dissipation of magnetic helicity, by its spatial redistribution
across the equator through turbulent fluxes driven by the hemispheric
distribution of kinetic helicity, and by the tentative presence
in the tangled dynamical magnetic field of plasmoids typical of
reconnection at large $Rm$.

\end{abstract}

\maketitle

\paragraph{Introduction.}
Magnetic fields pervading astrophysical fluid systems such as
stars and galaxies are commonly thought to be excited and further
sustained by a variety of self-amplifying
magnetohydrodynamic (MHD) dynamo
effects converting kinetic energy of turbulent flows
of electrically-conducting fluid into magnetic energy
\citep{moffatt78,branden05,shukurov07,roberts13,charbonneau14,brun17,rincon19}.
In order to sustain \textit{large-scale} magnetic fields via
a turbulent dynamo, however, some underlying system-scale
symmetry-breaking is generically required. Typically, this is provided
in astrophysical
systems by large-scale rotation and/or shear. In particular, by
breaking the parity/mirror invariance of an otherwise isotropic,
homogeneous turbulence, rotation makes the turbulence helical.
This creates the conditions for statistical dynamo effects
that can in principle amplify large-scale
magnetic fields exponentially on a rotation timescale
\citep{steenbeck66,raedler69a,raedler69b,moffatt82}. The most
well-known, the $\alpha$ effect \citep{parker55}, is generally
considered a key ingredient of magnetic-field generation
in the Sun.

While a helical turbulent dynamo provides an appealing phenomenological
explanation for the large-scale magnetism of rotating astrophysical
bodies, there remain major open questions regarding its actual
viability and efficiency in the regime of
large magnetic Reynolds numbers $Rm$ and comparable flow turnover
times and
correlation times, an astrophysically-relevant 
non-perturbative limit for which no analytical theory is
available \citep{moffatt70,krause80}
($Rm=UL/\eta$ is larger than $10^6$ in the Sun, and 
$10^{20}$ in galaxies ; $U$ denotes a typical
velocity field amplitude, $L$ is the typical scale of
the turbulence, and $\eta$ is the magnetic diffusivity).
Numerical studies have shown that large-scale exponential dynamo growth
driven by helical turbulence, such as rotating convection, is possible
at mild $Rm<O(100)$ (\cite{meneguzzi81,branden01}, see also
\cite{charbonneau14,brun17,branden15} for
reviews in various astrophysical contexts), however
this regime is still far from asymptotic in practice.

First of all, a distinct small-scale ``fluctuation'' dynamo
mechanism is activated beyond $Rm=O(100)$, that amplifies magnetic
fields on fast time and spatial scales comparable to the flow turnover scales 
\citep{kazantsev67,zeldovich84,schekochihin04,haugen04,iskakov07}.
This dynamo populates a large-$Rm$ turbulent MHD
fluid with dynamical small-scale fields affecting the structure of the
flow much faster than the helical dynamo can grow a large-scale
field \cite{kulsrud92,boldyrev01,boldyrev05}.
Besides, independently of a small-scale
dynamo, turbulent tangling of a growing large-scale field produces
dynamical magnetic fluctuations at increasingly smaller-scales
as $Rm$ increases (typically $\propto Rm^{-1/2}$), resulting in small-scale 
dynamical feedback on the flow well before the large-scale field
has itself saturated \citep{cattaneo96b}. In the case of homogeneous helical
turbulence producing an $\alpha$ effect, this problem takes a
particular pathological form: the large-scale field does
ultimately reach super-equipartition levels, but it can
only do so on a long, system-scale resistive time, a
consequence of a resistive bottleneck in the dissipation of
small-scale magnetic twists also responsible for the dynamical
reduction of the $\alpha$ effect \citep{boozer93,branden01,bhat16a,bhat21}. This is
usually referred to as the catastrophic $\alpha$-quenching
problem. Finally, the resistive-scale dynamics of saturated MHD
dynamos may undergo a fast-reconnection transition at $Rm=O(S_c)$,
where $S_c=O(10^4)$ is the critical value of the Lundquist number
$S=LV_A/\eta$ at which MHD reconnection becomes fast
\citep{loureiro07,uzdensky10,dong18} (assuming an Alfv\'en speed $V_A\sim
U$ in the saturated regime). Its implications for the
dynamics of helical fields, such as produced by the $\alpha$ effect, have so
far barely been touched on
\citep{eyink11,eyink11b,moffatt15,moffatt16,rincon19,cattaneo20,schekochihin21}.

We aim to further explore the nonlinear helical
dynamo problem at large $Rm$. A long-envisioned possible solution
to catastrophic quenching is via removal or spatial redistribution of
small-scale magnetic helicity by helicity fluxes
\citep{ji99,blackman00,kleeorin00,ji02,vishniac01,branden02,hubbard11,hubbard12,branden19}.
The most studied case \citep{branden01c,mitra11,delsordo13}
involves expulsion of magnetic helicity through system
boundary winds. Simulations up to $Rm\simeq 10^3$ suggest that a
regime with subdominant resistive effects is achieved at large $Rm$
\citep{delsordo13}. Alternatively, a similar
 state may be achieved via magnetic helicity fluxes driven through an
 equator by a hemispheric distribution of kinetic helicity,
 a simple configuration typical of rotating astrophysical systems 
\citep{branden01d}. This case has so far only been studied
at low $Rm$ where resistive effects dominate over helicity fluxes
\citep{mitra10a,mitra10b}. The numerical identification,
up to $Rm\simeq 3\times 10^3$, of a nonlinear helical state with
subdominant resistive effects is the main result of this work.

\paragraph{Model.}
We address the problem from a standard perspective of
magnetic helicity $\vA\cdot\vB$  dynamics ($\vA(\vec{r},t)$ is the magnetic
vector potential, $\vB=\curl\vA$ the magnetic field),
a local evolution equation of which can be derived from the induction equation,
\begin{equation}
\label{eq:helicitynonideal}
  \dpart{}{t}(\vA\cdot\vB)+\div{\vec{F}_{\mathcal{H}_m}} = -2\eta\,(\curl{\vB})\cdot\vB~,
\end{equation}
where
$\vec{F}_{\mathcal{H}_m} = c\left(\varphi\vB+\vE\times\vA\right)$
is a magnetic-helicity flux, $c$ is the speed of light,
$\vE$ is the electric field, and $\varphi$ is the electrostatic
potential. We split each field into a mean, large-scale part, defined below
as an average over the $(x,y)$ plane and denoted by an overline,
and a fluctuating, small-scale part, denoted by lower case letters,
$\vB(\vec{r},t)=\meanvB(z,t)+\fluctvB(\vec{r},t)$.
Manipulating the small- and large-scale components of the
induction equation, using $\vEMF\cdot\vB=0$, where $\vEMF=\fluctvU\times\vB$ is
the electromotive force (EMF) for a flow $\fluctvU$, one
obtains helicity budget equations
\begin{eqnarray}
  \label{eq:sshelicity}
\!\!\!\!\!\!  \dpart{}{t}(\mean{\fluctvA\cdot\fluctvB})+\div{\mean{\vec{F}}_{\mathcal{H}_{m,\mathrm{SS}}}}  &=& -2\,\meanvEMF\cdot\meanvB-2\eta\,\mean{(\curl{\fluctvB})\cdot\fluctvB}\,,\\
  \label{eq:lshelicity}
\!\!\!\!\!\!\dpart{}{t}(\meanvA\cdot\meanvB)+\div{\mean{\vec{F}}_{\mathcal{H}_{m,\mathrm{LS}}}}
  & = & 2\,\meanvEMF\cdot\meanvB-2\eta\left(\curl{\meanvB}\right)\cdot\meanvB\;.
\end{eqnarray}
In these equations, the first r.h.s.~terms describe the production of magnetic
  helicity, the second r.h.s.~terms its destruction by
  resistivity, and the second l.h.s.~terms describe the
  transport of magnetic helicity through the divergence of mean fluxes
  of fluctuating/mean helicities
\begin{eqnarray}
\mean{\vec{F}}_{\mathcal{H}_{m,\mathrm{SS}}} & = & c\left(\mean{\fluct{\varphi}\;\fluctvB}+\mean{\fluctvE\times\fluctvA}\right)~,\\
\mean{\vec{F}}_{\mathcal{H}_{m,\mathrm{LS}}} & = & c\left(\mean{\varphi}\,\meanvB+\meanvE\times\meanvA\right)
=\mean{\vec{F}}_{\mathcal{H}_{m}}-\mean{\vec{F}}_{\mathcal{H}_{m,\mathrm{SS}}}~,
\end{eqnarray}
where $\fluctvA$ and $\fluctvE$ denote fluctuations of the vector
potential and electric field.

\begin{figure}
   \includegraphics[width=\columnwidth]{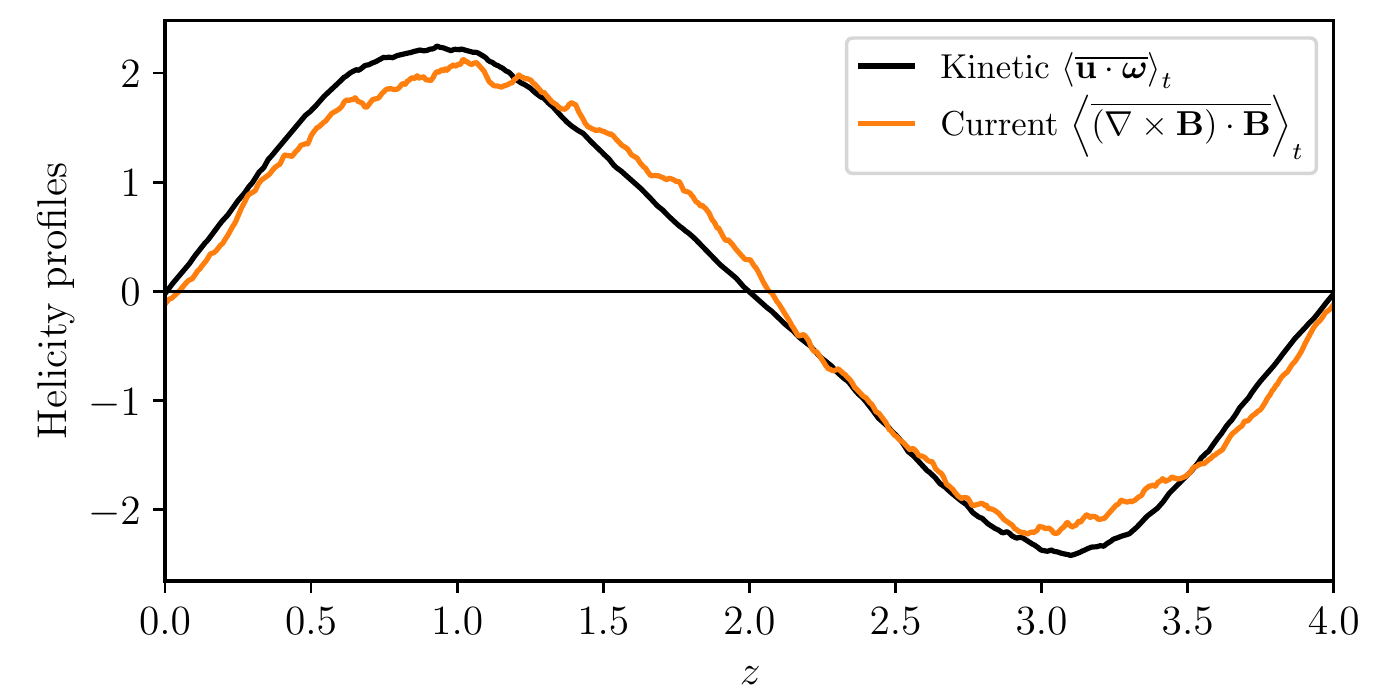}
\caption{Time- and $(x,y)$-averaged kinetic and current helicity $z$-profiles in a
  typical simulation (run T06, $Rm\simeq 2800$, $Re\simeq 700$, $L_f=1$,
  $L_z=4$).\label{kinetichelicityprofile}}
 \end{figure}

We compute these budgets in the Coulomb gauge $\div{\vec{A}}=0$
\citep{arlt01} for three-dimensional, spatially-periodic cartesian
simulations of nonlinear, helical, incompressible, viscous, resistive
MHD, carried out with the SNOOPY spectral code with 2/3 dealiasing \citep{lesur07}.
An inhomogeneous body force inspired by the Galloway-Proctor 
flow \citep{galloway92,tobias13} is implemented in the momentum
equation,
\begin{equation}
  \label{force}
  \begin{array}{l}
    \vec{f}(\vec{r},t)=k_f\,A_f\,\times \smallskip\\
\left(
    \begin{array}{c}
      \displaystyle{-2\sin\left(\f{2\pi y}{L_f}+\sin\omega_f t\right)\sin\f{2\pi z}{L_z}} \\
      \displaystyle{-2\cos\left(\f{2\pi x}{L_f}+\cos\omega_f t\right)\sin\f{2\pi z}{L_z}} \\
      \displaystyle{\sin\left(\f{2\pi x}{L_f}+\cos\omega_f
      t\right)+\cos\left(\f{2\pi y}{L_f}+\sin\omega_f t\right)}
      \end{array}
    \right)~, \end{array}
          \end{equation}
where $\omega_f$ and $A_f$ are a forcing frequency and amplitude,
$L_x=L_y\equiv L_f$  and $k_f=2\pi/L_f$ is the forcing wavenumber.
 This forcing drives a flow with a statistically-steady
   sinusoidal kinetic helicity profile in
   $z$. Fig.~\ref{kinetichelicityprofile} shows kinetic and
   current helicity profiles at saturation. Both are positive
   (resp. negative) for $z<L_z/2$ (resp. $z>L_z/2$), and
   change sign at ``equators'' $z=L_z/2$ and $z=0$ (replicated at
   $z=4$ in a periodic set-up). The mean, domain-averaged helicities are
   zero. Hence, this configuration i) mimicks a hemispheric distribution of
 kinetic helicity; ii) can potentially bypass catastrophic
 resistive quenching present in the standard homogeneous case by  enabling
 equatorial turbulent magnetic helicity fluxes;
 iii) keeps the system complexity minimal so as to maximise $Rm$.

 \begin{figure*}
  \includegraphics[width=2\columnwidth]{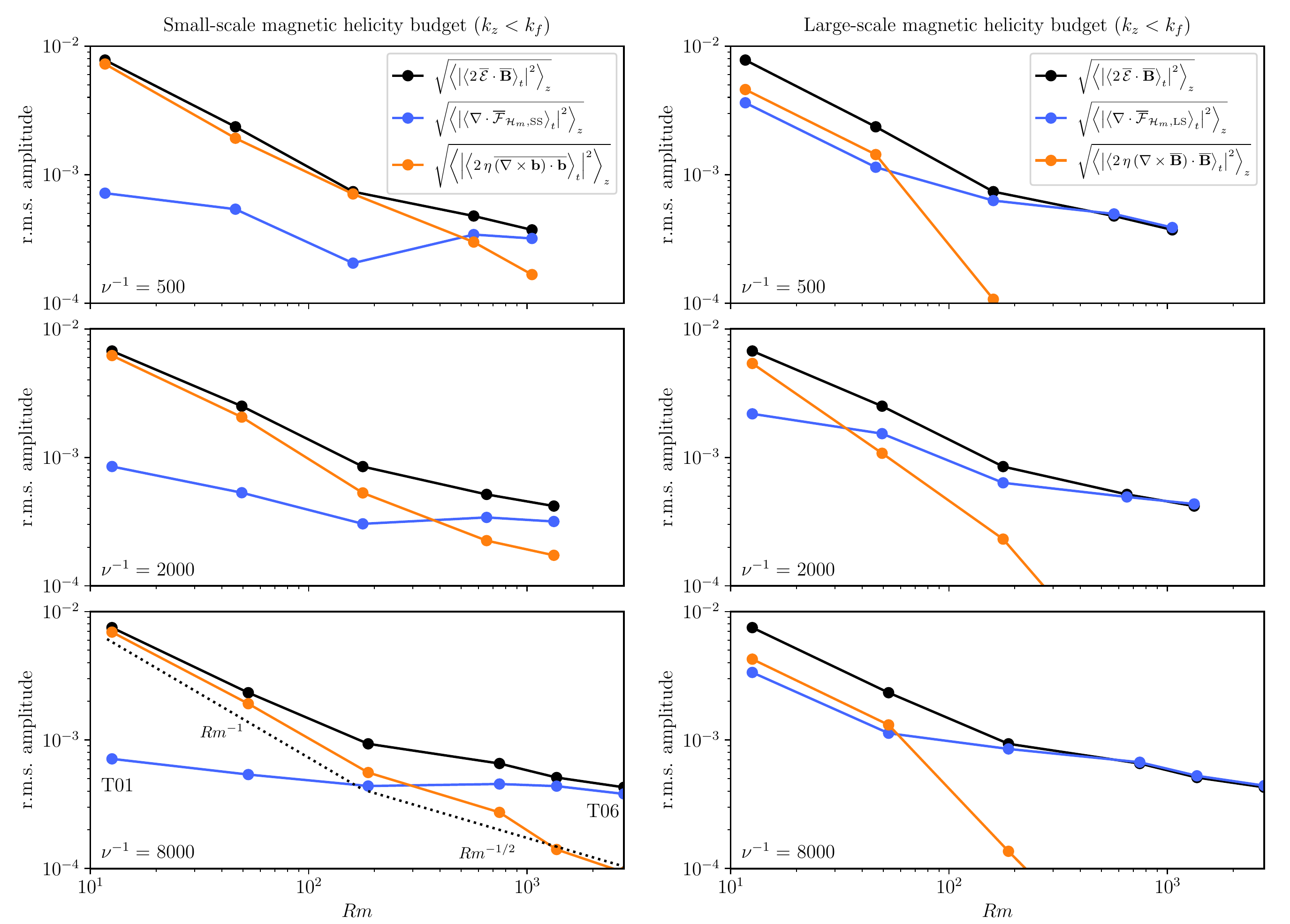}
  \caption{
Small- and large-scale helicity budgets on large scales in $z$ ($k_z<k_f$)
as a function of $Rm$, for different $Re$.\label{helicitybudgets}}
\end{figure*}

\renewcommand{\arraystretch}{1.1}
\setlength{\tabcolsep}{2.35pt}
\begin{table}
\begin{tabular}{lcccccccc}
  \hline\hline
   Run & $N_{(x,y)},N_z$ & $\nu^{-1}$ & $\eta^{-1}$ & $Re$ & $Rm$ & $u_\mathrm{rms}$ & $B_\mathrm{rms}$ & $\meanB_\mathrm{rms}$\\\hline
   V01  & $64^2,256$  & 500 & 125 & 46.7 & 11.7 & 0.59 & 0.49 & 0.39 \\
   V02  & $64^2,256$  & 500 & 500 & 46.2 & 46.2 & 0.58 & 0.57 & 0.42 \\
   V03  & $128^2,512$ & 500 & 2000 & 39.9 & 159.7 & 0.50 & 0.54 & 0.21 \\
   V04  & $128^2,512$ & 500 & 8000 & 35.6 & 570.0 & 0.44 & 0.58 & 0.14  \\
   V05  & $256^2,1024$ & 500 & 16000 & 32.9 & 1054.0 & 0.41 & 0.59 & 0.12 \\\hline
   M01  & $64^2,256$    & 2000 & 125  & 201.2 & 12.6 & 0.63 & 0.51 & 0.42 \\
   M02  & $64^2,256$    & 2000 & 500  & 197.6 & 49.4 & 0.62 & 0.56 & 0.38 \\
   M03  & $128^2,512$  & 2000 & 2000 & 177.0 & 177.0 & 0.56 & 0.59 & 0.32 \\
   M04  & $128^2,512$  & 2000 & 8000 & 163.3 & 653.2 & 0.51 & 0.60 & 0.18  \\
   M05  & $256^2,1024$ & 2000 & 16000 & 165.9 & 1327.5 & 0.52 & 0.60 & 0.12 \\\hline
   T01  & $128^2,512$  & 8000 & 125 & 804.4 & 12.6 & 0.63 & 0.48 & 0.37 \\
   T02  & $128^2,512$  & 8000 & 500 & 846.6 & 52.9 & 0.66 & 0.58 & 0.42 \\
   T03  & $128^2,512$  & 8000 & 2000 & 749.3 & 187.3 & 0.59 & 0.58 & 0.25 \\
   T04  & $128^2,512$  & 8000 & 8000 & 748.1 & 748.1 & 0.58 & 0.60 & 0.16 \\
   T05  & $256^2,1024$ & 8000 & 16000 & 683.4 & 1366.8 & 0.54 & 0.63 & 0.17 \\
   T06  & $512^2,2048$ & 8000 & 32000 & 694.5 & 2778.0 & 0.55 & 0.62 & 0.12 \\
  \hline\hline
\end{tabular}
\caption{Run index. $L_x,L_y=L_f=L_z/4$ for all runs.\label{tab1}}
\end{table}

We performed a parametric study for different Reynolds
$Re=u_\mathrm{rms}/(k_f\nu)$ and magnetic Reynolds numbers
$Rm=u_\mathrm{rms}/(k_f\eta)$, where $u_\mathrm{rms}$ is the
r.m.s.~flow amplitude (over time and space).
 We set $\omega_f=1$, $A_f=0.1$, $L_f=1$, and $L_z=4L_f$ in all
   simulations to ensure a minimal scale separation between the
   turbulence forcing scale and the scales of the helical inhomogeneity
   and emergent large-scale statistical dynamics.
Each run (Tab.~\ref{tab1}) was integrated for at least
50 forcing times $2\pi/\omega_f$ (175-200
flow turnover times $L_f/u_\mathrm{rms}$). To isolate the weaker
slow, large-scale signal from the fast turbulent noise, the magnitude
of each term in \equs{eq:sshelicity}{eq:lshelicity} was
estimated by Fourier-filtering them on $z$-scales larger
than the forcing scale ($k_z<k_f$), then taking the r.m.s.~values
(over $z$) of their time-averages after initial growth of the dynamo.

\paragraph{Results.}
Helicity bugdgets as a function of $Rm$ are
shown in Fig.~\ref{helicitybudgets}. 
The results are only weakly dependent on $Re$. At
low $Rm$, both budgets are characterized by a balance between
resistive and helical EMF terms. As $Rm$ reaches $50-100$,  the
large-scale budget transitions to a regime characterized by
a dominant balance between the large-scale $z$-flux of large-scale
helicity and EMF. However, for $Rm<300-500$, the dominant balance
  in the small-scale helicity budget remains
   between the resistive dissipation of small-scale helicity and
 EMF. Hence, the large-scale dynamics is still affected by
resistive effects in this $Rm$ range.
This regime is nevertheless interesting in that it can
only be realized for non-uniform flow helicity \citep{branden01d}.
It really takes $Rm>1000$ to reach a  regime characterized by a
  dominant non-resistive balance in both large- and small-scale
helicity budgets, the latter now being between the large-scale
$z-$flux of small-scale helicity and the EMF term. A weak
residual dependence of both terms on $Rm$ remains in the range of $Rm$ probed.

\begin{figure*}
\includegraphics[width=\columnwidth]{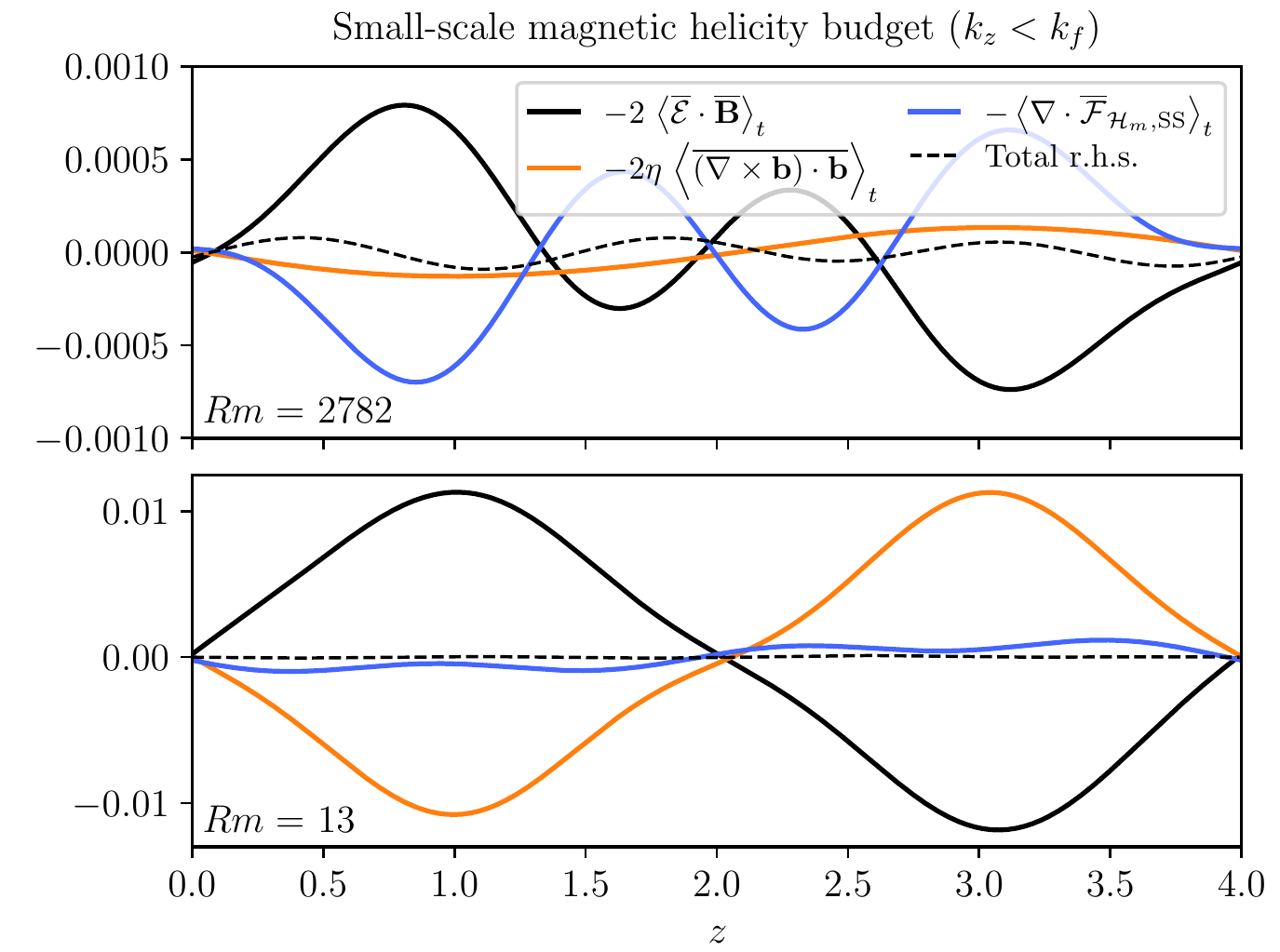}\includegraphics[width=\columnwidth]{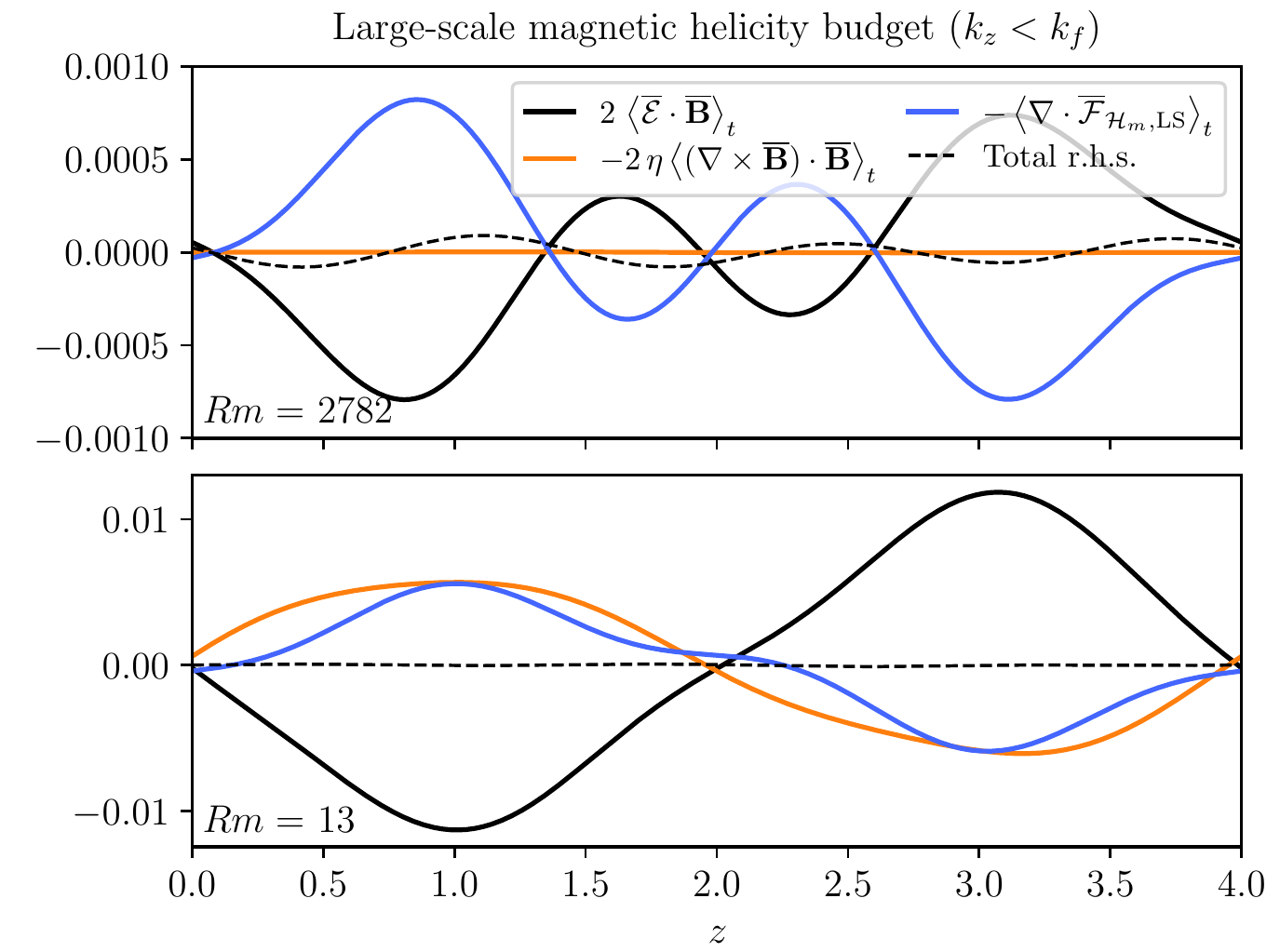}
\caption{Time-averaged helicity budgets on scales $k_z<k_f$ (T06:
  $Rm\simeq 2800$, $Re\simeq 700$ ; T01: $Rm\simeq 15$, $Re\simeq 800$).\label{profiles}}
\end{figure*}

A detailed comparison, between low and large-$Rm$ runs with identical
viscosity, of the time-averaged $z$-dependent quantities in
\equs{eq:sshelicity}{eq:lshelicity}, again filtered on $z$-scales
larger than the flow forcing scale, is shown in Fig.~\ref{profiles} to
make more explicit the transition between
the resistively-dominated and asymptotic regimes. Estimates of
  the flux divergences carried out in the Coulomb gauge
  were always found to be in good agreement with the calculation of
  the (gauge-independent) r.h.s. of
  \equs{eq:sshelicity}{eq:lshelicity} at large $Rm$, as expected in a
  statistically steady state \citep{delsordo13}. Hence, we are
  confident that the main trends reported here do not depend on
  our gauge choice.

\begin{figure}
  \includegraphics[width=\columnwidth]{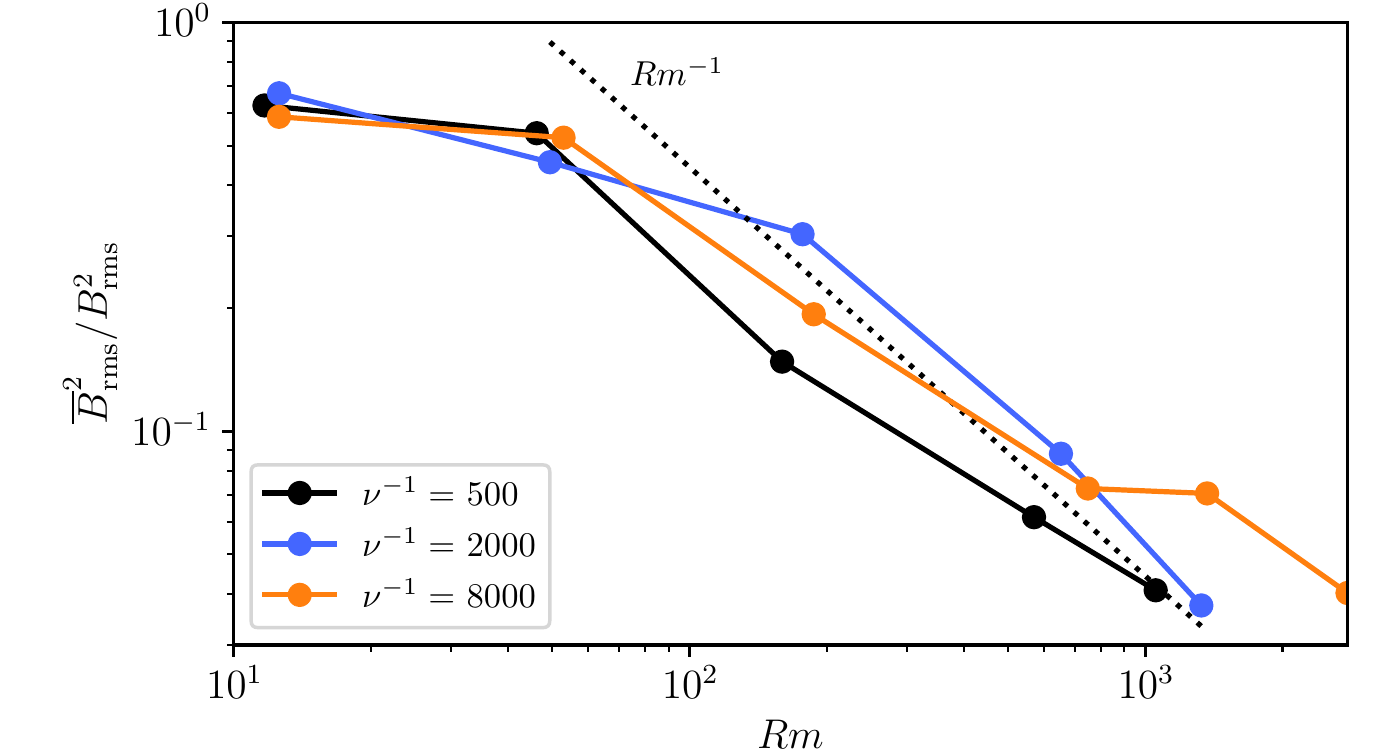}
  \caption{Time-averaged energy of $\meanvB$ relative to the total
    magnetic energy, as a function of $Rm$.\label{meanfieldamplitude}}
\end{figure}

Fig.~\ref{meanfieldamplitude} shows the energy of
$\meanvB$ as a function of $Rm$.
The results at intermediate $Rm$ are consistent with
a $Rm^{-1}$ scaling, in line with theory expectations and
earlier simulations \citep{branden09d,mitra10a}.
There is as yet no clear-cut evidence for an asymptotic
   regime entirely independent of $Rm$
(maybe because the small-scale helicity
dissipation term only seemingly decreases slowly as $Rm^{-1/2}$ at
large $Rm$), however we observe a clear deviation away of the
$Rm^{-1}$  scaling for the energy of the mean-field at the largest
$Re$ and $Rm$ probed. Mean-field models assuming turbulent diffusive
expressions for the helicity fluxes also suggest that convergence of
$\meanB^2_\mathrm{rms}/B^2_\mathrm{rms}$ towards an 
$Rm$-independent value
should be slow at large $Rm$ \citep{branden09d,mitra10b}.

\begin{figure}[h]
\includegraphics[width=\columnwidth]{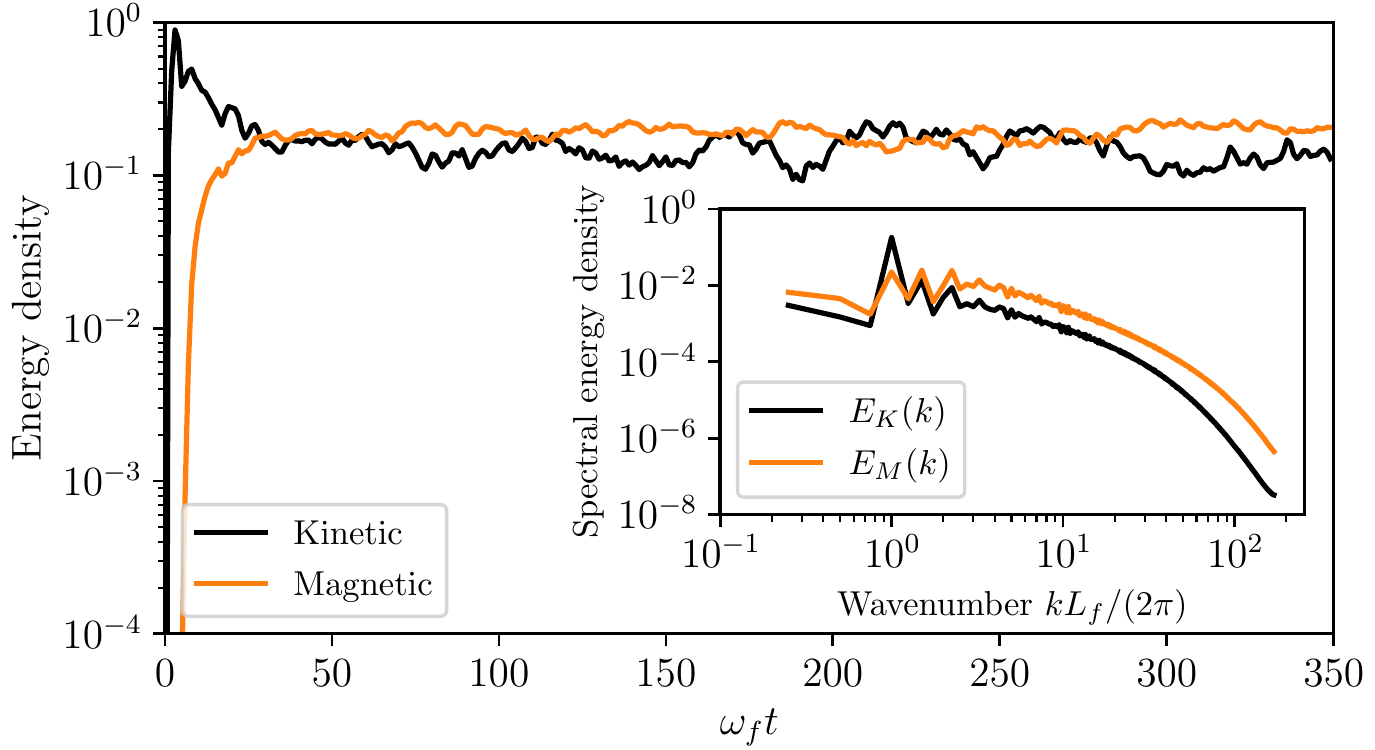}
\caption{Evolution of kinetic and magnetic energy den\-sities
    (T06: $Rm\simeq 2800$, $Re\simeq 700$). Inset: associated energy
    spectra in the saturated phase.\label{T06timevarspectra}}
\end{figure}

\begin{figure}[h]
\includegraphics[width=\columnwidth]{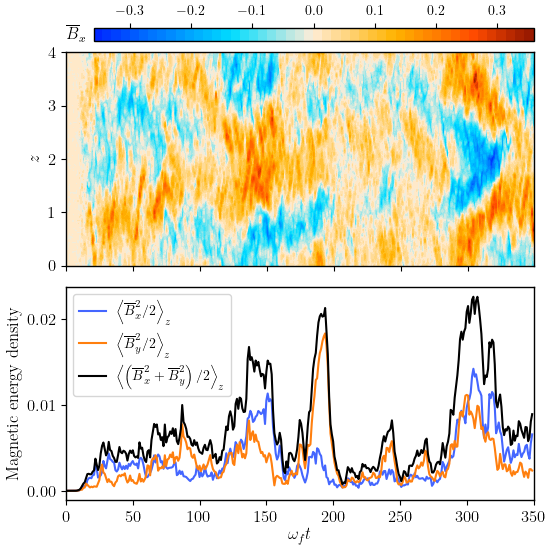}
\caption{Evolution of (top) $\meanB_x(t,z)$ and
  (bottom) energy density of mean field (T06: $Rm\simeq 2800$,
   $Re\simeq 700$).\label{T06time}}
\end{figure}

\begin{figure}
\includegraphics[width=\columnwidth]{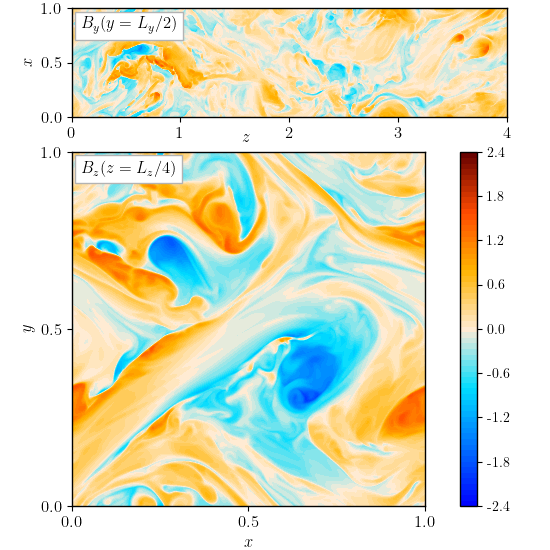}
\caption{$(x,z)$ (top) and $(x,y)$ (bottom) 
  out-of-plane magnetic-field snapshots (T06: $Rm\simeq 2800$,
  $Rm=700$).\label{T06snap}}
\end{figure}

For our parameters, convincing access to a regime with subdominant
resistive contributions required
a (spectral) resolution of 512 per $L_f$ (run T06, $Rm\simeq 2800$, $Re\simeq
700$). The evolution of energy densities, and time-averaged energy
spectra in the statistically steady state of T06 are
shown in Fig.~\ref{T06timevarspectra}.
The evolutions of the large-scale
field component $\meanB_{x}(z,t)$ and large-scale magnetic energy
densities are shown in Fig.~\ref{T06time}.
$\meanvB$ displays bursty oscillations, on a timescale $\sim
50-60\,\omega_f^{-1}$ ($\sim 30$ turnover times), that appear to
propagate spatially towards the
  equator associated with the node of kinetic helicity at $z=2$,
a likely consequence of the symmetry-breaking flow helicity profile in
$z$ \citep{branden09d}. Snapshots at the end of the run
(Fig.~\ref{T06snap}) show complex, turbulent
  magnetic structures with tentative nascent plasmoids
typical of reconnection in nonlinear tangled magnetic fields
at large $Rm$ \citep{rincon19,schekochihin21}, e.~g. at
  $(x,z)\simeq (0.5,0.75);(0.75,1.3);(0.8,3.2);(0.75,3.5)$.
The ``small-scale'' horizontal structure
visible in the bottom plot is the direct imprint of the forcing at
    $L_f$ and is distinct from the weaker, but larger-scale emergent
    statistical order in $z$ visible in \fig{T06time}.

\paragraph{Discussion.}
Numerical results showing a similar transition in the presence of
advective ``wind'' boundary losses of helicity have been
obtained previously \citep{delsordo13}, albeit with a larger
transition $Rm$ (their simulations at lower $Rm$ with no wind also
hinted at solutions involving
turbulent diffusive fluxes, but with different symmetries).
In all cases, the strong dependence of the saturated state on
$\eta$ is circumvented at large $Rm$ by non-resistive helicity
fluxes. The nonlinear state achieved
here, anticipated in \citep{branden09d,mitra10a,mitra10b},
is particularly appealing in that it stems from a very simple
inhomogeneous, hemispheric distribution of kinetic helicity also
typical of rotating astrophysical systems. While the
solution is dominated by small-scale fields, a clear magnetic activity
pattern migrating towards the equator is present on
scales larger than the flow forcing scale. Further investigations
are needed to determine whether a mean shear
may boost the generation of a streamwise ``azimuthal''
large-scale field, and how such a shear, or the transition to the
low $Pm$, large $Rm$ regime typical of stellar dynamos may affect
the properties of the identified travelling wave pattern.

Modelling helicity fluxes as turbulent diffusive fluxes also suggests
that the transition $Rm$ (to the regime with a subdominant resistive term)
scales as $(k_f/\overline{k})^2$,
where $\overline{k}$, the scale of $\meanvB$, should be comparable to the
helicity modulation scale \citep{mitra10b}. If this scaling applies,
something we could unfortunately not test due to limited computing
resources, the asymptotic regime of large-scale astrophysical dynamos
typically involving large scale-separations may be at significantly
higher $Rm$ than that determined here for $L_z/L_f=4$. As global
simulations are currently limited to $Rm$ of a few
hundreds (also uncomfortably close to the small-scale dynamo
threshold), this raises the question of their lack of
asymptoticity for the foreseeable future. Our results may provide a useful
reference point to assess such future simulations in this respect.

An in-depth understanding of this large-$Rm$ MHD state
remains to be developed. One may be tempted to interpret it
``classically'' as the nonlinear outcome of an $\alpha$ effect dynamo
\citep[see also \cite{ji99} for theoretical work
  involving helicity fluxes]{parker55,steenbeck66, moffatt78}.
Oscillations of a weak large-scale field on top of helical turbulent
MHD background also suggest a (possibly connected)
phenomenological interpretation in terms of simple
large-scale magnetoelastic waves in small-scale tangled fields
\citep{hosking20}. Finally, while we may have tentatively observed
  reconnection plasmoids in these simulations, providing a new
  independent estimate of the minimal (spectral) resolution required
  to accomodate fast reconnection in turbulent MHD, much more
  numerical work will be required in the future at even higher resolution
  to fully characterize it, and its so far poorly-understood
  posible effects on large-scale magnetic field generation at
  asymptotically large $Rm$.

\begin{acknowledgments}
\paragraph{Acknowledgements.}
I thank Alexander Schekochihin, Jonathan Squire and
Nuno Loureiro for many stimulating discussions. This work was granted
access to the HPC resources of CALMIP under allocation 2019-P09112 and
IDRIS under GENCI allocation 2020-A0080411406.
\end{acknowledgments}

\bibliography{dynamo}

\end{document}